\documentclass[aps,prl,twocolumn,floatfix]{revtex4}
\usepackage{graphicx,graphics,epsf,psfig,epsfig}
\usepackage{subfigure}

\def\beq{\begin{equation}}
\def\eeq{\end{equation}}
\def\beqa{\begin{eqnarray}}
\def\eeqa{\end{eqnarray}}

\begin{document}

\title{Lattice tube model of proteins}

\author{Jayanth R. Banavar$^1$, Marek Cieplak$^2$,
 and Amos Maritan$^3$}

\address{
$^1$104 Davey Laboratory, Department of Physics,
The Pennsylvania State University,
University Park, Pennsylvania 16802\\
$^2$Institute of Physics, Polish Academy of Sciences,
Al. Lotnik{\'o}w 32/46, 02-668 Warsaw, Poland \\
$^3$Universit{\'a} degli Studi di Padova, Dipartimento di Fisica
and INFN, via
Mazzolo 8, 35100, Padova, Italy }




\begin{abstract}

{\small
We present a new lattice model for proteins that incorporates
a tube-like anisotropy by introducing a preference for mutually parallel
alignments in the conformations. The model is demonstrated to
capture many aspects of real proteins. 
}

\end{abstract}

\maketitle

\vspace*{-0.8cm}
\hspace*{1.5cm} PACS Numbers: 87.15.He, 87.15.Cc, 87.15.Aa

\vspace*{0.5cm}


There have been several physics-based attempts to distil the
essential features of the protein problem and notable success in
capturing many of the key ingredients has been achieved using
lattice models \cite{Dillreview}.  Such coarse-grained
descriptions allow a virtually exact analysis of many properties
and provide a useful framework for understanding experimental
results. Indeed, valuable progress has been made within the
simplified description of a lattice model with just two types of
amino acids denoted by $H$ and $P$ representing hydrophobic and
polar behaviors. The principal theme of this letter is to present a
new lattice model of proteins, which takes into account a
previously overlooked key attribute of chain molecules -- the
context of amino acids within a chain. We benchmark the behavior
of this model with the well-studied HP lattice model and show that
the new model faithfully captures several attributes of real
proteins.


There are clear hints,
manifested by the many common characteristics of proteins
\cite{Banavar}, that proteins may be simpler than one might
expect. Protein structures are constructed in a modular manner
from common building blocks -- helices, hairpins and sheets
connected together by tight turns. Further, the total number of
distinct protein folds seems to be of the order of just a few
thousand \cite{Chothia}.

The simplest model of an unconstrained object is a hard sphere. A
collection of hard spheres exhibits both fluid and crystalline
phases on changing the volume fraction.  When objects are tethered
together in the form of a chain, it is no longer appropriate to
consider them as spheres. There is a special direction that one
may associate with each object which is tangent to the chain and
is defined by the adjoining particles along the chain. It is
therefore more appropriate to model the objects making up a chain
by means of discs or coins, for which the heads-to-tails direction
is distinct from the two other directions. This picture of
tethered coins leads to a tube-like description of a chain
molecule \cite{Banavar}. Just as symmetry plays a key role in
determining the nature of ordering of unconstrained particles (the
phases associated with a collection of spheres are vastly simpler
than the liquid crystal phases of anisotropic objects), the
anisotropy inherent in a tube leads to new behavior.  Recent work
\cite{Banavar} has shown that the tube picture can be used to
understand the conventional polymer phases and the novel phase of
matter used by Nature to house protein native state structures in
a unified way and for the development of a framework for
understanding the common character of proteins.

There are three key features of a tube description that one ought
to incorporate in a lattice model: self-intersections of a tube
are not allowed, the local radius of curvature of a tube can be no
smaller than the tube radius 
and in a compact state, there is a tendency
for nearby tube segments to be parallel (indeed both helices and
sheets have tube segments alongside and parallel to each other
leading to a cooperative placement of hydrogen bonds \cite{Liwo}).
The first two features are built into a model of a self-avoiding
chain on a lattice.  Our focus here is on considering the effects
of introducing the third. 

In order to illustrate the key idea, we will consider a 16 amino
acid (aa) self-avoiding chain on a square lattice. There have been
numerous previous studies \cite{Dillreview} of this system within
the standard HP model context and its generalizations \cite{Rios}.
In the standard $HP$ model, one
ascribes a favorable energy $-1$ for a $HH$ contact (two H aa
which are {\em not} next to each other in sequence but sit next to
each other in the lattice) and zero energy otherwise. Here, in
addition we pay attention to the context that the contact occurs
in. Figure 1 illustrates three distinct types of contacts (denoted
by an index $m$) depending on the degree to which the segments
containing the aa in contact are parallel to each other. The
energy assigned to a $HH$ contact of type $m$ in the Tube HP (THP)
model is denoted by $e_m$. In what follows, let us choose $e_m$ to
be $-3$, $-2$, and $-1$ for $m$=3, 2, and 1 respectively thereby
favoring the parallelism of nearby segments.  In the standard HP
model $e_m= -1$ independent of $m$.

In order to understand the role of sequence heterogeneity, it is
useful to consider a generalized model in which the energies are
described by
\begin{equation}
E \;=\; \sum_{i<j} \; e_m \; \Delta (i-j,m) \; [ \delta_{i,H} \delta_{j,H}
\;+\;(1 - \epsilon ) \; D_{ij}] \;\;,
\end{equation}
where
\begin{equation}
D_{ij}\;=\;(\delta_{i,H} \delta_{j,P} + \delta_{i,P} \delta_{j,H}
+\delta_{i,P} \delta_{j,P} ) \;= \; 1\;-\;\delta_{i,H}\delta_{j,H}\;\;.
\end{equation}
Here, $\Delta (i-j,m)$ is equal to 1 if the amino acids $i$ and
$j$ form a contact and 0 otherwise. When such a contact exists,
the energy of attraction associated with it depends on the index
$m$. $\delta _{i,H}$ is defined to be equal to 1 if amino acid $i$
is hydrophobic and 0 if it is polar. Similarly, $\delta_{i,P} = 1
- \delta_{i,H}$ is equal to 1 if amino acid $i$ is polar.
Depending on the choice of the $e_m$ parameters, one obtains the
HP or THP models. The limiting cases correspond to $\epsilon =1$,
i.e. the 'standard' THP or HP models, and $\epsilon =0$ -- the
case of a homopolymer made of $H$ amino acids.

For the 16-aa chain, all sequences and all possible conformations
can be enumerated exactly. There are interesting differences in
the energy landscape of the HP and the THP models.  One may
determine the sequences which have a unique ground state and the
number of distinct designable conformations, which house these
sequences, as a function of $\epsilon$ (see inset of Figure 2).
For a homopolymer ($\epsilon $=0), the HP model has no designable
structure -- all compact conformations are degenerate and have the
same energy. Thus in the absence of sequence specificity,
there is no pre-selection of protein-like structures among
compact conformations.
Thus in the absence of sequence specificity, there is
no protein-like behavior. When a weak heterogeneity is introduced
by turning on a small $\epsilon$, the HP energy landscape becomes
rugged and each of the 69 maximally compact conformations become
designable but with a weak thermodynamic stability.  Thus the
funnel-like energy landscape \cite{funnel} arises only on turning
on the full degree of sequence heterogeneity.

\begin{figure}
\epsfxsize=3.6in \centerline{\epsffile{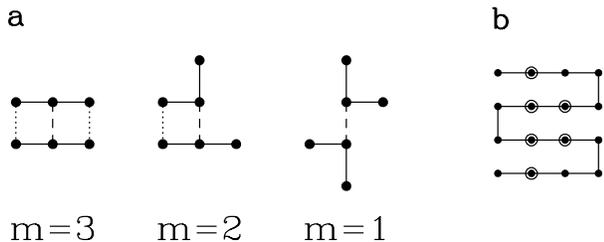}}
\vspace*{-3.5cm}
\caption{ {\small
Panel a: Sketch of three contact environments in the THP model. The dashed
line denotes a contact. Panel b:  The optimal
structure for the THP model. The circled 
represent the hydrophobic core and have H aa in them
more than 87 \% of the time for the sequences that fold into this
conformation when $\epsilon = 1$.
 }}
\end{figure}

This is in sharp contrast to the behavior of the THP model --
here, even for a homopolymer, one obtains a unique ground state,
akin to either a helix or a sheet in two dimensions
(see Figure 1), selected not by considerations of the chemistry of
the aa but rather by the overarching principles of geometry and
symmetry. Interestingly, in the limit of small $\epsilon$, all
$2^{16}$=65536 sequences have a unique ground state in the THP 
model and none in the HP model.
When $\epsilon =1$, one obtains 10579 and 1539
designable sequences in the THP and HP models respectively (see
Figure 2) folding into 684 and 456 distinct folded structures.
Furthermore, the number of sequences folding into the most
designable structure are 637 and 26 for the two models.

\begin{figure}
\epsfxsize=3.7in \centerline{\epsffile{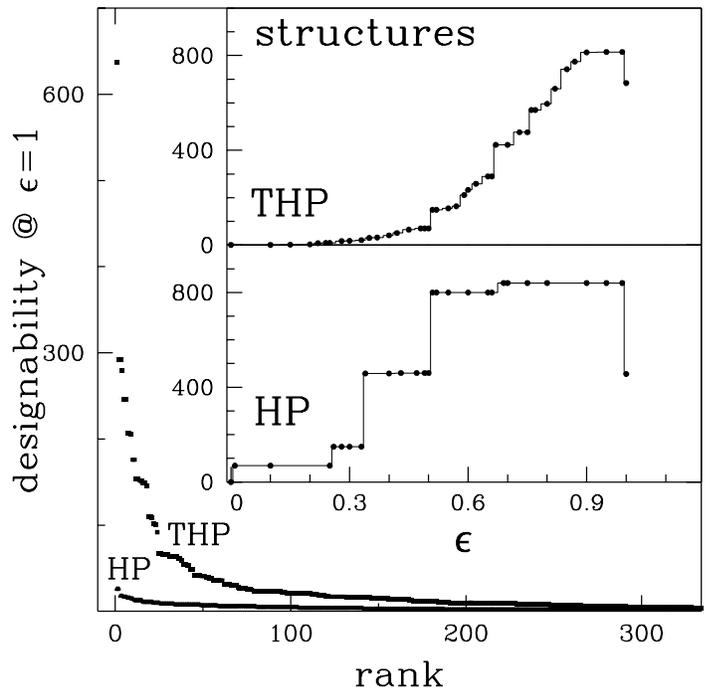}}
\caption{ {\small
Rank ordered values of the number of sequences that fold into the
given structure for all of the designable structures at $\epsilon
=1$. The inset shows the number of designable structures as a
function of $\epsilon$ (see text).
}}
\end{figure}

The thermodynamic stability of a sequence is characterized by the
folding transition temperature, $T_f$, at which the equilibrium
probability of being in the native state conformation is equal to
$\frac{1}{2}$. The spread in the values of $T_f$ is nearly three
times larger in the THP model than in the HP case. 
The most stable THP sequence  folds into the structure
shown in Figure 1b, whereas the most stable HP sequence
folds to a structure which is not maximally compact. In order to
describe the folding kinetics, we take a sequence at a temperature
equal to its $T_f$ value and consider 10 batches of 101
trajectories and determine the first passage time to the native
state starting from an unfolded conformation. The time evolution
\cite{monte} is a Monte Carlo process which satisfies detailed
balance. The kinetic moves consist of single bead moves (the kink
flips and rotations of the terminal segments) with probability 0.2
and of two bead ``crankshaft" moves with probability 0.8.  A
median folding time is determined for each batch and averaged over
all batches to yield a measure of the folding time, $t_{fold}$.
Our calculations were carried out for 12 sequences in each model
(the top 10 sequences in $T_f$ values and the sequences ranked 20
and 30). In all cases, the THP model exhibits more rapid 
two-state folding
than the HP model with the ratio of the folding times for the 12
sequences ranging between 0.10 and 0.47.

The framework of evolution in life works through both the DNA
molecule and the functionally useful protein molecule. Mutations
of the DNA molecule lead to the possibility of new proteins, whose
selection, in turn, leads to an enhancement of the number of such
DNA molecules. As pointed out by Maynard-Smith\cite{Maynard}, as
the sequence undergoes mutation, there must be a continuous
network that the mutated sequences can traverse without passing
through any intermediaries that are non-functioning.  Thus, one
seeks a connected network in sequence space for evolution by
natural selection to occur.  There is considerable evidence that
much of evolution is neutral \cite{Kimura}. 

We have investigated the topology of connections \cite{Chan} between 
the designable structures resulting from point mutations in the sequence
(the change of one aa from H to P or vice versa).
Indeed, while one has a ``random walk" in sequence space
that forms a connected network,
there is no similar continuous variation in structure space.
When $\epsilon = 1$, 39.3 \% or 605
of the HP sequences do not belong to 
the connected network envisioned by Maynard-Smith in sharp
contrast to the THP model for which only 13 of the 10579 sequences,
i.e. 0.12 \%, do not belong to the network. 
The THP model is vastly better connected than the HP model,
as illustrated in Figure 3. The former exhibits
approximate scale-free behavior \cite{Reka} while the latter is
more akin to a random network with low mean coordination number
(Figure 4).

\begin{figure}
\epsfxsize=3.7in \centerline{\epsffile{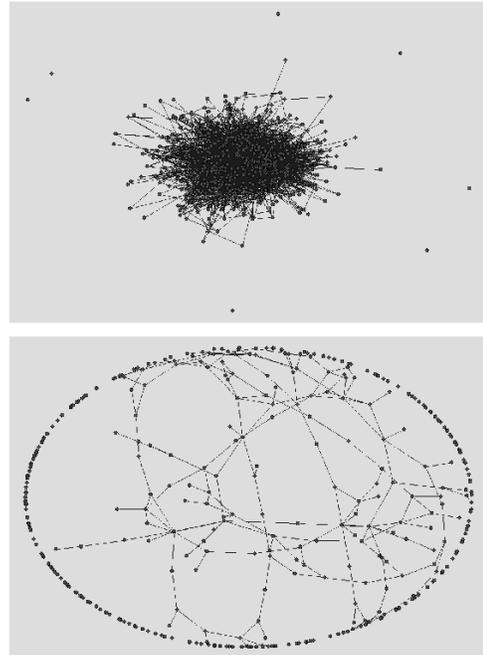}}
\vspace*{-1cm}
\caption{ {\small
Network topologies (using Pajek)
of designable structures resulting from
point mutations in the sequence. The top and bottom panels are for the
THP and HP models respectively.  
 }}
\end{figure}

\begin{figure}
\epsfxsize=3.9in \centerline{\epsffile{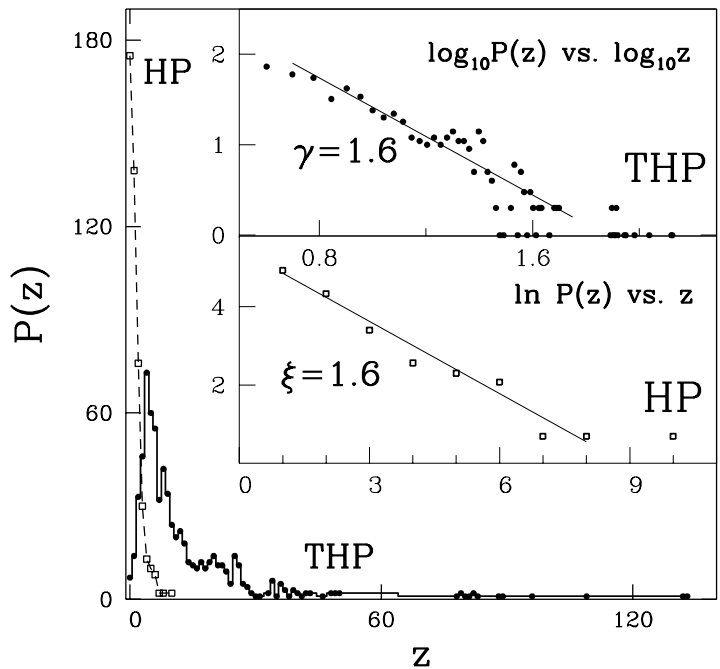}}
\caption{ {\small
Probability distribution, $P(z)$, of the effective coordination  
number for the network of designable structures shown in Fig. 3.
The inset is a plot of the same data in
a log-log scale (the top panel) for the THP model and in a
log-linear scale for the HP model.
The results illustrate the approximate validity of
$P(z) \sim z^{-\gamma}$ and  $P(z)\sim exp{-z/\xi}$
for the THP and HP models  respectively.
 }}
\end{figure}

In summary, we find that the tube lattice model captures many of
the key characteristics of protein behavior in a superior way
compared to conventional lattice models. The key advantage of
studying a tube on a lattice compared to a more realistic
continuum analysis \cite{Banavar} is that one can often carry out
an exact analysis for short chains and obtain insights on real
protein behavior. As an illustration, we conclude with a simple
analysis of a few hundred proteins \cite{Dima} to determine the
propensity of amino acid pairs in contact \cite{Tsai} to be in
specific environments characterized by the $m$-index introduced
above. Specifically, we look at the type of contact between aa $k$
and aa $l$ along the sequence and categorize it in the following
manner: the specific aa pair involved in the contact, their
sequence separation $s =  \mid k-l \mid$ equal to 2, 3, 4 or
greater than 4 and the number of contacts $m$ between the two
groups of aa ($k-1,k,k+1$) and ($l-1,l,l+1$) which can range
between 1 and 9. (The geometry of the lattice model in two
dimensions allow for only three values $1$, $2$ or $3$  of the contact
environment index $m$.) We have determined
\begin{equation}
\chi _2(k,l,s,m)\;=\;
\frac{[n(k,l,s,m)\;-\;p(k,l,s,m)]^2}{p(k,l,s,m)} \;\;.
\end{equation}
Here $n$ is the number of contacts and $p$ the expected number of
contacts based on chance: $p(k,l,s,m)= a q(k,l,s)$, where $q$ is
the number of the specific aa  pairs at distance $s$ and 
$a=\sum_{kl} n(k,l,s,m) / \sum_{kl} q(k,l,s)$.
A large value of $\chi_2$ corresponds to
a strong signal that aa $k$ and aa $l$ prefer to make or avoid a
contact in the environment defined by the $s$ and $m$ indices
(Table 1) and would be useful in the development of scoring
functions for protein structure recognition \cite{Dima}.

The tube idea reveals a deep underlying simplicity in the protein problem.
In standard approaches, the sequence of amino acids is believed to play
a key role in sculpting the free energy landscape and determining
its native state structure. Here, instead, the free energy landscape
is sculpted predominantly by symmetry and geometry and the sequence
plays a vital role in the {\em selection}
of the native state from a predetermined
menu.  Unlike sequences and functionalities, which are shaped by
the powerful forces of evolution, the menu of putative native state
structures is immutable and is determined by physical law.  Indeed,
this fixed backdrop provides the initial basis for selection in molecular
evolution. DNA which make proteins that are able to fold readily into
one of the predetermined folds pass the initial screening. 
An additional level of filtering completes the selection process of
proteins that are not only good folders but are also able to interact well
with ligands and other proteins and play a useful functional role.

We are grateful to Istvan Albert for helpful advice. This work was
supported by KBN (grant 2 P03B 032 25), COFIN MURST 2003, INFM,
NASA, NSF IGERT grant DGE-9987589 and the NSF MRSEC at Penn State.


\newpage



\centerline{TABLE CAPTION}

{\bf Table I}.   {\small
The list of aa pairs with $\chi _2 (k,l,s,m)$ larger than
65. In the ensemble of proteins that were studied, 
there are 97 918, 97 525, 97132, and 17 506 983 aa pairs
with $s$ equal to 2, 3, 4, and gretaer than 4 respectively.
336 110 of the pairs form contacts: 
22.1, 11.9, 10.4, and 55.6 \%  of them correspond
to $s$=2, 3, 4, and $> 4$ respectively. For each $s$, the distribution
of the contacts over the contact type $m$ is uneven. For $s$=2 and 3,
most of the contacts, 36.5 and 59.8 \% respectively, 
corresponds to $m$=8.
These contacts typically correspond to interactions within helices.
Amino acids with long and/or forked side groups (L, K, Q, R, E) are 
more likely to form local contacts with a large $m$.
On the other hand, the smallest amino
acid, G, is much less likely to form such contacts, as evidenced
by the aversion in the pairs G-G, G-P, and G-S for $s$=2 and $m$=8. 
The propensity of aa A to participate in short range contacts with
a large $m$ (also for $s$=4) is also due to its size: A is small
enough to allow for participation in conformational twists, but it
is sufficiently big to facilitate formation of many contacts.
For $s$=4, 67.3 \% of the contacts ocupy $m$=6. Finally, for $s > 4$,
45.4 \% of the contacts occupy $m$=1 and 2 almost equally.
These contacts usually correspond to links between 
secondary structures, e.g. between two helices or between
a helix and a turn, through a pair of hydrophobic amino acids which
are unlikely to be a G. The C-C covalent attraction results in
non-local contacts over a range of $m$ values.
}

\newpage

\hspace*{3.5cm} Table I

\begin{tabbing}
\=xxxxxxxxxxxxxxxx\=xxxxxxxxxxxxxxxxxxxx\=xxxxxxxx\=xxxx \kill

\> {\bf \underline{ aa pairs}} \> {\bf \underline{ attraction/aversion}}
\> {\bf $\;$\underline{s}} \>{\bf\underline{m}} \\

\> {\bf  V-I} \>   \hspace*{0.9cm}attraction   \>     2 \>  4  \\
\> {\bf AL-AEQKR} \>  \hspace*{0.9cm}attraction  \>      2  \> 8 \\
\> {\bf G-PSG }  \>  \hspace*{1.7cm}aversion
\>  2  \> 8  \\
\vspace*{-0.3cm}
\> ------------------------- \> ------------------------------ \>
------------ \> ------ \\
\> {\bf  A-AQIR $\;$ L-ALQ }\> \hspace*{0.9cm}attraction  \>    3  \> 8   \\
\vspace*{-0.3cm}
\> ------------------------- \> ------------------------------ \>
------------ \> ------ \\
\> {\bf  A-A $\;$ L-LA $\;$ E-R}\>\hspace*{0.9cm}attraction  \>  4 \> 6   \\
\> {\bf  G-V} \> \hspace*{1.7cm}aversion  \>  4 \> 6  \\
\vspace*{-0.3cm}
\> ------------------------- \> ------------------------------ \>
------------ \> ------ \\
\> {\bf L-IFVLMWY } \>  \hspace*{0.9cm}attraction  \>   $>$4 \> 1  \\
\> {\bf V-IFVMW $\;$ F-FWY} \> \hspace*{0.9cm}attraction  \>   $>$4 \> 1   \\
\> {\bf I-FIWM $\;$ C-C $\;$ M-FY} \> \hspace*{0.9cm}attraction \>
  $>$4 \> 1 \\
\> {\bf A-G $\;$ G-DST } \> \hspace*{1.7cm}aversion \>  $>$4 \> 1 \\

\> {\bf L-IFVLMWY $\;$W-Y} \>  \hspace*{0.9cm}attraction  \>   $>$4 \> 2  \\
\> {\bf VI-IF $\;$ F-FW $\;$ C-C}\>\hspace*{0.9cm}attraction\>$>$4 \> 2   \\
\> {\bf L-LF $\;$ I-V $\;$C-C }\> \hspace*{0.9cm}attraction \> $>$4 \> 3 \\
\> {\bf C-C }\> \hspace*{0.9cm}attraction \> $>$4 \> 4 \\

\> {\bf V-VI $\;$ I-I } \> \hspace*{0.9cm}attraction  \>  $>$4 \>5   \\
\> {\bf V-LVIFT $\;$ I-I }   \> \hspace*{0.9cm}attraction \>  $>$4 \> 6   \\
\vspace*{-0.3cm}

\end{tabbing}

\end{document}